\documentclass[11pt]{article}

\newcommand{\paperauthor}{D\"{u}zg\"{u}n K\"{u}\c{c}\"{u}k}
\newcommand{\paperaffiliation}{Department of Digital Forensics Engineering, F\i{}rat University, Elaz\i\u{g}, T\"{u}rkiye}
\newcommand{\paperorcid}{0000-0003-4937-9237}

\usepackage[T1]{fontenc}
\usepackage{lmodern}
\usepackage[a4paper,margin=2.5cm]{geometry}
\usepackage{graphicx}
\usepackage{booktabs}
\usepackage{tabularx}
\usepackage{xltabular}
\usepackage{array}
\usepackage{amsmath}
\usepackage{caption}
\usepackage{float}
\usepackage{placeins}

\usepackage{xurl}
\usepackage[hidelinks]{hyperref}
\usepackage{orcidlink}
\hypersetup{pdftitle={Forensic Analysis of the HackRF One with PortaPack Mayhem Firmware},pdfauthor={D\"{u}zg\"{u}n K\"{u}\c{c}\"{u}k}}

\newcolumntype{L}{>{\raggedright\arraybackslash}X}
\newcolumntype{P}[1]{>{\raggedright\arraybackslash}p{#1}}

\newcommand{\figfile}[1]{%
  \begingroup\setlength{\fboxsep}{1.5pt}\setlength{\fboxrule}{0.5pt}%
  \sbox0{#1}%
  \fbox{\resizebox{\dimexpr\wd0-2\fboxsep-2\fboxrule\relax}{!}{\usebox0}}\endgroup}

\title{Forensic Analysis of the HackRF One with PortaPack Mayhem Firmware}
\author{\paperauthor\,\orcidlink{\paperorcid}\\[2pt]
  \small \paperaffiliation}
\date{11 October 2026}

\begin{document}

\maketitle

\begin{abstract}
Software-defined radios (SDRs) have become an inexpensive, portable platform for replay attacks, location spoofing and jamming. The HackRF One can operate stand-alone when fitted with a PortaPack interface board running the open-source Mayhem firmware. No peer-reviewed study has addressed the forensic analysis of this device family. This work establishes, at source-code level, where forensically relevant data reside on the microSD card of a device running Mayhem v2.0.1, in which format, when they are written and how they should be interpreted. The methodology is applied to an SD card image that the author acquired from their own device after four months of real use, and every inference is drawn from the image alone. The workflow covers timestamp rules that depend on the real-time clock (RTC) battery state, traces of on-device trimming and text editing, cluster-level recovery of deleted in-phase/quadrature (IQ) captures, and decoding of recorded signals cross-checked with an independent decoder. In the example image, the final part of a deleted IQ capture and a trimmed-away capture segment are recovered, and six captures are attributed to a single rolling-code on-off keying (OOK) transmitter sharing one 32-bit identifier. The study provides practitioners with an artifact catalogue, interpretation rules and acquisition recommendations.
\end{abstract}


\section{Introduction}\label{sec:intro}

Software-defined radios (SDRs) are devices that use software in place of most of the circuitry that processes radio signals. As a result, a single piece of hardware can work as both a receiver and a transmitter over a wide frequency range. The HackRF One is an open-hardware, low-cost SDR that operates from 1~MHz to 6~GHz in half-duplex mode~\cite{hackrfdoc}. In this mode, it either receives or transmits at any one time. When the PortaPack interface board and the Mayhem firmware~\cite{mayhem} are added, the device becomes a stand-alone tool, independent of a computer, with its own touchscreen, battery and microSD (Secure Digital) card. This tool can record radio signals digitally, as in-phase/quadrature (IQ) samples, and retransmit them. It can also generate fake Global Positioning System (GPS) signals, Automatic Dependent Surveillance--Broadcast (ADS-B) messages, with which aircraft report their positions, and Bluetooth Low Energy (BLE) packets. These capabilities can be used in offences such as location spoofing, jamming communications and the replay attack, in which the signal of a car or garage remote control is recorded and retransmitted later.

Kouwen et al.~\cite{kouwen2018}, in their study of the forensic examination of two-way radio equipment, stated that SDR hardware capable of record-and-replay, such as the HackRF One, requires the ``special attention'' of law-enforcement experts and called for research on such hardware. Established methods exist for the forensic analysis of embedded systems and Internet of Things (IoT) devices such as drones~\cite{clark2017}--\cite{stankovic2021}\nocite{lee2023}, smart speakers~\cite{chung2017}, game consoles~\cite{eichhorn2024}--\cite{read2024}\nocite{barrsmith2021,pessolano2019}, smartwatches~\cite{gregorio2019}, \cite{odom2019} and vehicle infotainment systems~\cite{shin2022}, \cite{marques2026}. However, no comparable study was found for SDRs. The closest studies either use an SDR as a signal-collection tool~\cite{binazhar2021} or address the file signatures of a different radio tool, the Flipper Zero~\cite{fereli2025}. In digital forensics, the files or traces that a device or software leaves behind during use and that provide information about user activity are called artifacts. This study addresses the artifacts left by the HackRF One/\allowbreak PortaPack.

This study answers the following research questions:
\begin{enumerate}
  \item Where, and in what format, are forensically relevant data stored on a HackRF One/\allowbreak PortaPack running Mayhem v2.0.1, and which component writes them and when?
  \item How can user activity, namely which applications were used, on which frequencies and when, be reconstructed from these artifacts? How should the timestamps be interpreted?
  \item How can deleted, trimmed or edited data be detected and recovered?
  \item How can the type of recorded radio signals be determined, their contents decoded, and this decoding validated?
\end{enumerate}

The contributions of the study are as follows:
\begin{enumerate}
  \item For Mayhem v2.0.1, an SD card artifact catalogue in which every artifact is linked to the source-code file that writes it, and a mapping of the application settings files to the applications (Section~\ref{sec:catalogue}).
  \item An explanation of how file dates are formed through the \texttt{DATE\_FILEFLAG} file when the backup battery that powers the device's real-time clock (RTC) is absent, and an inference of the time zone that exploits the fact that the device, Windows and Linux write file times according to different rules (Section~\ref{sec:time}).
  \item Detection of trimming with IQ Trim and text editing with Notepad on the device; recovery of deleted and trimmed captures from the free space of the card (Section~\ref{sec:deleted}).
  \item A workflow for decoding the remote-control signals in the captures, tested on the Mayhem sample captures by comparison with the independent decoder \texttt{rtl\_433}. This workflow determines whether a signal is sent with on-off keying (OOK), by which rule the bits are converted into pulses, and whether different captures come from the same transmitter (Sections~\ref{sec:iq} and~\ref{sec:validation}).
\end{enumerate}

The method was applied to an SD card image taken from the author's own device and containing real, unscripted usage data; all inferences were drawn from the traces in the image alone, and the image is publicly available~\cite{hackrfimage}. These contributions enable an examiner who encounters a HackRF One/\allowbreak PortaPack to determine, with rules based on the source code, which file to look at, how to read the timestamps and how to recover and decode deleted captures.

The rest of the paper is organized as follows. Section~\ref{sec:related} describes related work, Section~\ref{sec:device} the device and the Mayhem data model, and Section~\ref{sec:methods} the materials and methods. Section~\ref{sec:artifacts} presents the artifacts and interpretation rules on the example image. Section~\ref{sec:validation} contains the validation, Section~\ref{sec:discussion} the discussion, Section~\ref{sec:limitations} the limitations and Section~\ref{sec:conclusion} the conclusions.

\FloatBarrier
\section{Related Work}\label{sec:related}

Related studies were identified through Crossref, OpenAlex, arXiv and general web searches using combinations of the terms ``HackRF'', ``PortaPack'', ``software-defined radio'' and ``forensic'', and by following the reference lists and citations of the papers found. No peer-reviewed study performing a forensic analysis of the HackRF One or PortaPack/Mayhem was found.

Clark et al.~\cite{clark2017} decoded the flight records stored in the internal memory of the DJI Phantom III drone and on its microSD card, which uses the FAT32 (32-bit File Allocation Table) file system, and published an open-source parser named DRone Open source Parser (DROP). They resolved the record format by reverse engineering, recovered deleted records only in part, and validated their results with a second tool and with hash values. Chung et al.~\cite{chung2017} proposed an approach that combines cloud, client and device artifacts in the Amazon Alexa ecosystem. Eichhorn et al.~\cite{eichhorn2024} performed predefined actions in sequence on a Steam Deck and, by comparing the images taken after each action, linked every trace to the action that created it. In this way, they interpreted the files and configuration on the internal solid-state drive of the device, which runs Linux, and also identified deleted data. Studies on the Nintendo Switch~\cite{barrsmith2021}, the 3DS~\cite{pessolano2019}, \cite{read2024}, smartwatches~\cite{gregorio2019}, \cite{odom2019}, vehicle infotainment systems~\cite{shin2022}, \cite{marques2026} and DJI remote controllers~\cite{lee2023} follow a similar structure. These studies first identify the device and its version, then acquire the data, resolve the location and format of the artifacts and validate the results. The present study follows the same structure, but none of these studies addresses an SDR or recorded radio signals.

Studies on messaging applications~\cite{anglano2014}--\cite{walnycky2015}\nocite{anglano2016,anglano2017,jaeckel2025} determine the location of artifacts and the meaning of database fields with controlled test data and present the results as generalizable interpretation rules. Anglano et al.~\cite{anglano2017} also proposed a workflow for the analysis of any Android application. Jaeckel et al.~\cite{jaeckel2025}, in turn, used the open source code of the application directly when interpreting artifacts. In this study, the fact that Mayhem is open source is used for the same purpose.

Few studies are close to radio devices. Kouwen et al.~\cite{kouwen2018} proposed an examination process for two-way radio equipment in which the device memory and the programming software are read with the vendor's own tool. They tested the process on scenarios and pointed to SDR hardware as a future research area. Their process does not address deleted data or the content of radio signals. Bin Azhar and Abadia~\cite{binazhar2021} used an SDR not as an object of examination but as a tool for collecting radio signals for forensic purposes, and evaluated their approach with experiments. They did not examine the data that the SDR itself stores. Fereli et al.~\cite{fereli2025} examined the microSD card of the Flipper Zero, developed file signatures for its file types and recovered deleted files with these signatures. They validated the signatures with recovery tests. None of these studies examines both the files stored by a radio device and the signals recorded in those files.

The present study differs from these studies in three respects. First, the format and meaning of each artifact were derived from the firmware source code instead of reverse engineering or a vendor tool. Second, deleted and trimmed data were detected and recovered at the level of FAT32 clusters, not only by file signatures. Third, the recorded IQ captures themselves were analysed, and the OOK signals in them were demodulated. The results were validated against the source code, with a second tool and on sample captures.

\FloatBarrier
\section{Device and Data Model}\label{sec:device}

\begin{figure}[!htb]
\centering
\figfile{\includegraphics[width=\textwidth]{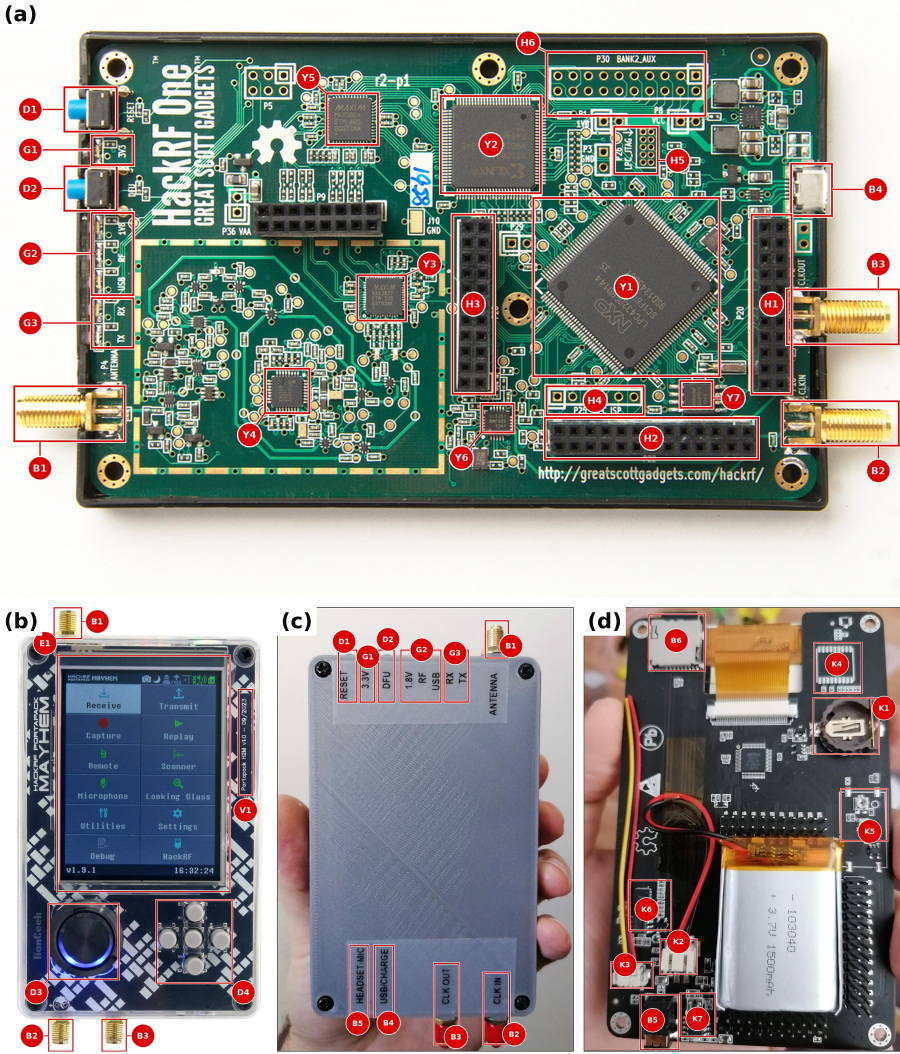}}
\caption{Hardware components of the HackRF One and PortaPack: (a) HackRF One main board, (b) PortaPack H2M front face, (c) PortaPack H2 enclosure edges, (d) PortaPack H2 inner face. Image sources: (a) Great Scott Gadgets, HackRF repository (GPL-2.0)~\cite{hackrfdoc}; (b)--(d) Mayhem firmware wiki~\cite{mayhemwiki}. Codes and markings were added in this study. The source images are not covered by the licence of this paper and remain subject to their original terms.}
\label{fig:components}
\end{figure}

At the heart of the HackRF One is the NXP LPC4320 microcontroller, which runs the firmware, that is, the device software. This chip has no non-volatile memory; at every boot the firmware is loaded from a separate flash memory connected through the Serial Peripheral Interface (SPI)~\cite{hackrfdoc}. The circuits that receive and transmit the radio signal consist of the MAX2837 transceiver, the RFFC5072 mixer and the 8-bit MAX5864 analog-to-digital/digital-to-analog converter (ADC/DAC). The board carries an SMA (SubMiniature version A) antenna connector, CLKIN and CLKOUT ports for an external clock reference, a micro-USB (Universal Serial Bus) port, RESET and DFU (Device Firmware Upgrade) buttons, and six status light-emitting diodes (LEDs). The PortaPack plugs into the three expansion headers of the HackRF and adds a 3.2-inch touchscreen, a rotary encoder, direction keys, an audio circuit, a microSD slot and an RTC backup battery~\cite{mayhemwiki}. Figure~\ref{fig:components} shows these components. The code letter indicates the type of component: B port or connector, D button, G status light, E display, Y chip, H pin header, K PortaPack internal component, V revision marking. The same component carries the same code in all panels; the codes were chosen from letters that cannot be confused with the printed labels on the board. The original numbers in the source images have been covered with these codes. The screenshot in panel~(b) belongs to Mayhem v1.9.1, whereas the examined device runs v2.0.1. From a forensic point of view, the most important components are the microSD slot (B6), the RTC backup battery (K1), the SPI flash holding the firmware (Y7) and the micro-USB port (B4); the relationship of these components to the artifacts is discussed in Section~\ref{sec:discussion}.

Mayhem is a firmware built on the ChibiOS/RT 2.6.8 real-time operating system and distributed under the GNU General Public License (GPL)~\cite{mayhem}. The user interface and the signal processing run on two separate cores of the microcontroller. The firmware is written to the SPI flash. Since v2.0.0, some of the applications are kept on the SD card as \texttt{APPS/*.ppma} files. The device's own memory has no file system for user data; all user data that persist when the device is switched off are on the SD card.

\begin{figure}[!htb]
\centering
\figfile{\includegraphics[width=0.8\textwidth]{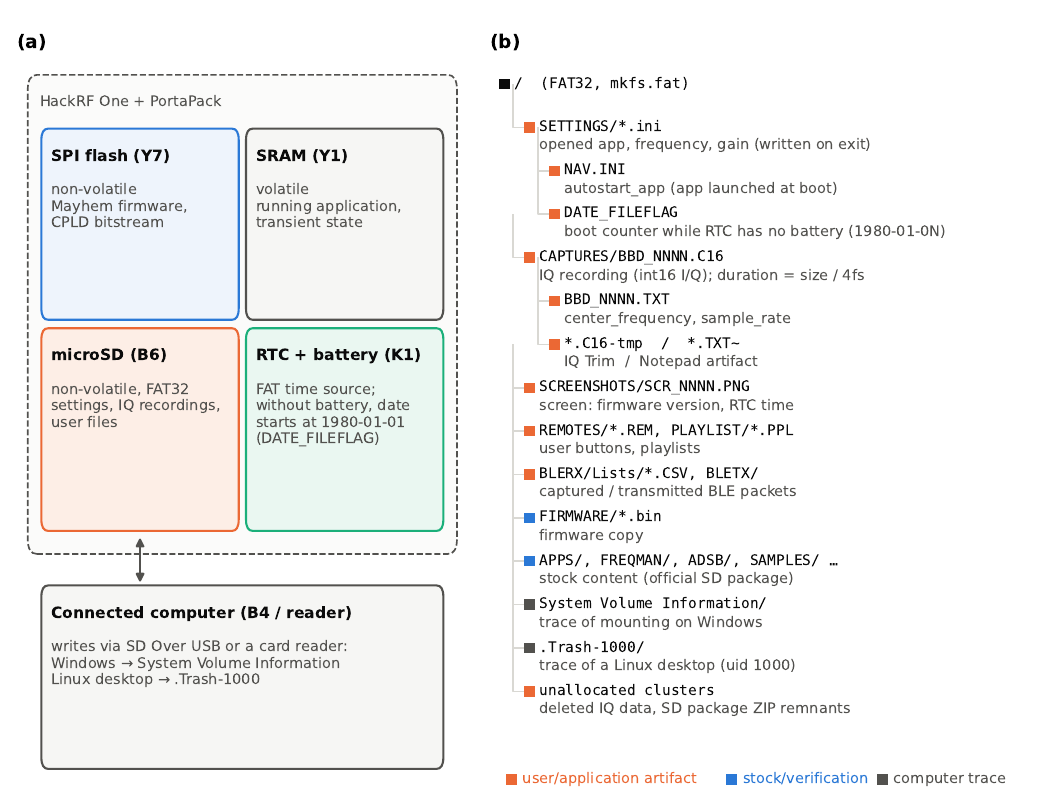}}
\caption{(a) Data locations on the device; the codes in parentheses are those of Figure~\ref{fig:components}. (b) Mayhem v2.0.1 SD card directory tree. CPLD: complex programmable logic device; SRAM: static random-access memory.}
\label{fig:artifactmap}
\end{figure}

Mayhem uses fixed directory names in the root directory of the SD card~\cite{mayhem}. Figure~\ref{fig:artifactmap}(a) summarizes the four data locations on the device and whether the data in these locations persist when the device is switched off, and Figure~\ref{fig:artifactmap}(b) summarizes the SD card directory tree and the forensic value of each directory. The ``unallocated clusters'' in Figure~\ref{fig:artifactmap}(b) are not a directory but the area that the file system marks as free and in which deleted data may be found. The Capture application writes radio captures to the \texttt{CAPTURES/} directory. Each application saves its settings to the \texttt{SETTINGS/} directory when it is closed. Screenshots are written to the \texttt{SCREENSHOTS/} directory. The stock directories, such as \texttt{APPS/}, \texttt{FREQMAN/}, \texttt{ADSB/} and \texttt{SAMPLES/}, come from the official SD card package; therefore, whether they have been modified can be verified by comparison with the official package. The computers into which the card is inserted also leave their own traces on the card.

\section{Materials and Methods}\label{sec:methods}

This section describes the examined device and data set, the integrity principles applied throughout the analysis, and the analysis methods used.

\subsection{Device and data set}\label{sec:dataset}

The examined device belongs to the author: a HackRF One with Mayhem v2.0.1 installed and a PortaPack attached to it. The device was used by the author during August--December 2024 without a predefined list of actions. The data set therefore reflects real, unscripted use, unlike the controlled test data in related studies~\cite{clark2017}, \cite{eichhorn2024}.

The image of the SD card, that is, an exact byte-for-byte copy of the card, was taken by the author. The card was removed from the device, inserted into another computer and copied in raw (dd) format with FTK Imager. The image is not a copy of the whole card but of the FAT32 partition that starts after the first 1~MiB of the card. This indicates that only this partition was selected in FTK Imager as a logical-volume source, rather than the whole card. The image was later stored as the single member of a ZIP archive. The local-time and Coordinated Universal Time (UTC) fields that the archive keeps for the image file show that the copying was completed at 2024-12-20 08:42:28~UTC (11:42:28 local time). The hash report produced by FTK Imager during copying was not kept. The earliest integrity evidence after acquisition is therefore the cyclic redundancy check (CRC-32) value that the ZIP archive recorded for the image, and the image matches this value. The hash values of the image were first computed at the start of the examination (2026-10-07) and verified against these values throughout the examination. A hash is the digital fingerprint of a file; if even a single byte of the file changes, the hash value changes completely. The Message-Digest~5 (MD5) and Secure Hash Algorithm~256 (SHA-256) hash algorithms are used in this study. The materials and tools are given in Table~\ref{tab:materials}. A third-party transmission picked up from the surroundings during capture is used only to demonstrate the method. This transmission is a BLE discovery packet, and the last two bytes of its identifier have been masked. The free text in the recorded files consists of the author's own entries. The 433~MHz remote control whose signals were captured belongs to the author. Transmissions made with the device were directed only at the author's own devices and were short-range. The settings files of the transmitting applications show only that these applications were opened; the paper makes no claim about the content or timing of any transmission.

\begin{table}[!htb]
\centering
\caption{Materials, tools and versions.}
\label{tab:materials}
\small
\begin{tabularx}{\textwidth}{@{}P{3.3cm}L@{}}
\toprule
Item & Details \\
\midrule
Device & HackRF One + PortaPack, Mayhem v2.0.1; hardware revision and serial number not recorded, SPI flash not read \\
Image & \texttt{hackrf.dd} (FAT32 partition), 3,908,042,752~B, SHA-256 \texttt{99f6f02e\ldots{}948e} \\
Acquisition tool & FTK Imager, raw (dd) format, logical volume source (no record of the version) \\
Firmware reference & Mayhem v2.0.1, commit \texttt{8e90c65a}, released 2024-04-07 04:37:29 UTC~\cite{mayhem} \\
Examination system & Ubuntu 26.04.1 LTS, Linux 7.0.0 \\
File system & The Sleuth Kit 4.12.1~\cite{sleuthkit}, dosfstools 4.2, a FAT32 parser written for this study \\
Integrity & hashdeep 4.4, sha256sum \\
Carving and archives & foremost 1.5.7, PhotoRec 7.2, binwalk 2.4.3, 7-Zip 26.00 \\
Signal & Python 3.14 (numpy 2.3.5, scipy 1.16.3, matplotlib 3.10.7), \texttt{rtl\_433} 25.12~\cite{rtl433}, multimon-ng 1.3.1~\cite{multimonng}, SoX 14.7 \\
\bottomrule
\end{tabularx}
\end{table}

\FloatBarrier
\subsection{Evidence-based inference, integrity and read-only access}\label{sec:evidence}

All inferences were drawn from the traces in the image alone, and every interpretation was checked against the Mayhem source code. The author's knowledge of how the device had been used was not used as evidence. This set-up is close to the situation of an examiner who is faced with a device whose usage history is unknown; however, because the same person used and examined the device, it is not a substitute for independent validation (Section~\ref{sec:limitations}). The image file was made read-only, was not mounted as a disk by the operating system, and all access was read-only~\cite{nist80086}, \cite{iso27037}. At the end of the examination, the hash values of the image were recomputed and verified to match the initial values.

\FloatBarrier
\subsection{Analysis}\label{sec:analysis}

The examination was carried out with a six-stage workflow. The input to the workflow is the SD card image (Section~\ref{sec:dataset}). Each stage uses the output of the previous one and answers a specific examination question:
\begin{enumerate}
  \item Acquisition and integrity: The basis of the analysis is showing that the evidence has not changed. Therefore, the hash values of the image are taken first, and all operations are performed on a read-only copy (Sections~\ref{sec:dataset} and~\ref{sec:evidence}).
  \item File-system analysis: FAT32 divides the card into equal-sized blocks called clusters and keeps a table of the clusters in which each file is stored. Which files are on the card and which have been deleted are derived from this structure. Unallocated space consists of the clusters marked as free. Because the data of a deleted file may remain there until new data are written over it, this area is also scanned (Sections~\ref{sec:firmware}, \ref{sec:catalogue} and~\ref{sec:deleted}).
  \item Artifact interpretation: The meaning of each file found is determined by locating the function in the Mayhem source code that writes that file. This stage runs together with the second stage; each file seen in the file system is immediately matched to the source code (Section~\ref{sec:catalogue}).
  \item Timeline: Because FAT32 timestamps carry no time-zone information, each timestamp is read according to the system that wrote it, namely the device, Windows or Linux. The state of the RTC battery is also taken into account (Section~\ref{sec:time}).
  \item IQ signal analysis: Whether the captures contain a signal and, if so, how it is encoded is determined. The decoded bits are validated by comparison with an independent decoder (Section~\ref{sec:iq}).
  \item Hardware mapping: Finally, each artifact is associated with the hardware component that produces it. This reveals the components that leave no trace on the SD card and must be examined separately (Section~\ref{sec:discussion}).
\end{enumerate}

At all stages, a chain-of-custody log, the tool versions, the scripts and the intermediate outputs were recorded; at the end of the examination, the hash values of the original files were verified again. Table~\ref{tab:workflow} summarizes the methods and tools used in these stages and the output of each stage.

\begin{table}[!htb]
\centering
\caption{Stages, methods and outputs of the forensic examination workflow.}
\label{tab:workflow}
\small
\begin{tabularx}{\textwidth}{@{}P{3.4cm}LP{3.4cm}@{}}
\toprule
Stage & Methods and tools & Output \\
\midrule
1. Acquisition and integrity & MD5 and SHA-256 hashes; read-only working copy & Verified copy \\
2. File-system analysis & \texttt{fsstat}, \texttt{fls}, \texttt{istat}, \texttt{fsck.fat -n}; deleted entries and unallocated space & File list, deleted entries \\
3. Artifact interpretation & Mapping to the Mayhem v2.0.1 source code (together with stage~2) & Artifact catalogue \\
4. Timeline & \texttt{mactime}; writing system, time zone and RTC state & Event timeline \\
5. IQ signal analysis & Envelope, power spectrum, spectrogram; OOK decoding and comparison with \texttt{rtl\_433} & Signal class, bit sequences \\
6. Hardware mapping & Association with the component codes in Figure~\ref{fig:components} & Component--artifact table \\
\bottomrule
\end{tabularx}
\end{table}

The details of the stages are as follows. The file system was listed with The Sleuth Kit~\cite{sleuthkit}, following established file-system forensic methods~\cite{carrier2005}. The same results were cross-checked with an independent parser written for this study according to the FAT32 specification~\cite{fat32spec}. Deleted file records, unallocated space and old folder records that are no longer linked to any folder were also scanned. Such old folder records are called orphan directory clusters. The meaning of each artifact was derived from the Mayhem v2.0.1 source code. This approach takes the place of the reverse engineering used in related studies~\cite{clark2017}, \cite{lee2023} and links every interpretation to the code that produces it. The timeline was generated from \texttt{mactime} and the file list, and the timestamps were interpreted according to the system that wrote them (Section~\ref{sec:time}). The IQ captures were examined in three ways: the envelope, the Welch power spectral density (PSD) and the spectrogram. The envelope shows how the signal power changes over time. The PSD shows how the power is distributed over frequencies. The spectrogram shows how the frequency content changes over time. OOK signals, that is, signals in which the transmitter sends information by switching on and off, were decoded as follows: the envelope was split into on and off segments with a threshold, the durations of these segments were grouped, and the bits were extracted from the durations. The result was compared with an independent decoder called \texttt{rtl\_433}. The main parameters are:
\begin{enumerate}
  \item Envelope: 50~$\mu$s moving average for the overview and 20~$\mu$s moving average for demodulation.
  \item Welch PSD: 4,096-point Hann window.
  \item Spectrogram: 1,024 points; a colour range of $-115\ldots{-35}$~dBFS/Hz common to all captures. dBFS denotes decibels relative to a full-scale complex sinusoid. When I and Q are clipped at their extreme values simultaneously, the envelope reaches about +3~dBFS.
  \item Packet detection: 10~dB above the noise floor.
  \item Pulse threshold: for each packet, the mean of the on level and the noise floor, with $\pm 2$~dB hysteresis.
\end{enumerate}

Generative artificial intelligence assistance was used in developing the analysis scripts; every result produced with this assistance was validated by the independent tool and source-code comparisons in Section~\ref{sec:validation}.

\FloatBarrier
\section{Artifacts and Interpretation Rules}\label{sec:artifacts}

In this section, each subsection corresponds to an examination question. Each subsection first gives the general rule and then shows how the rule applies to the example image.

\FloatBarrier
\subsection{Device and firmware identification}\label{sec:firmware}

The Mayhem version can be determined from three independent sources on the SD card. The first is the firmware file under \texttt{FIRMWARE/}; the hash value of this file is compared with the official release package. The second is the version text on the bottom line of the device's own screenshots. The third is the modification times of the stock SD package files; these times are compared with the release time of the version. The firmware that actually runs on the device is in the SPI flash and can only be read from the device, with \texttt{hackrf\_spiflash -r}. The SD card evidence is therefore indirect.

\begin{figure}[!htb]
\centering
\figfile{\includegraphics[width=0.35\textwidth]{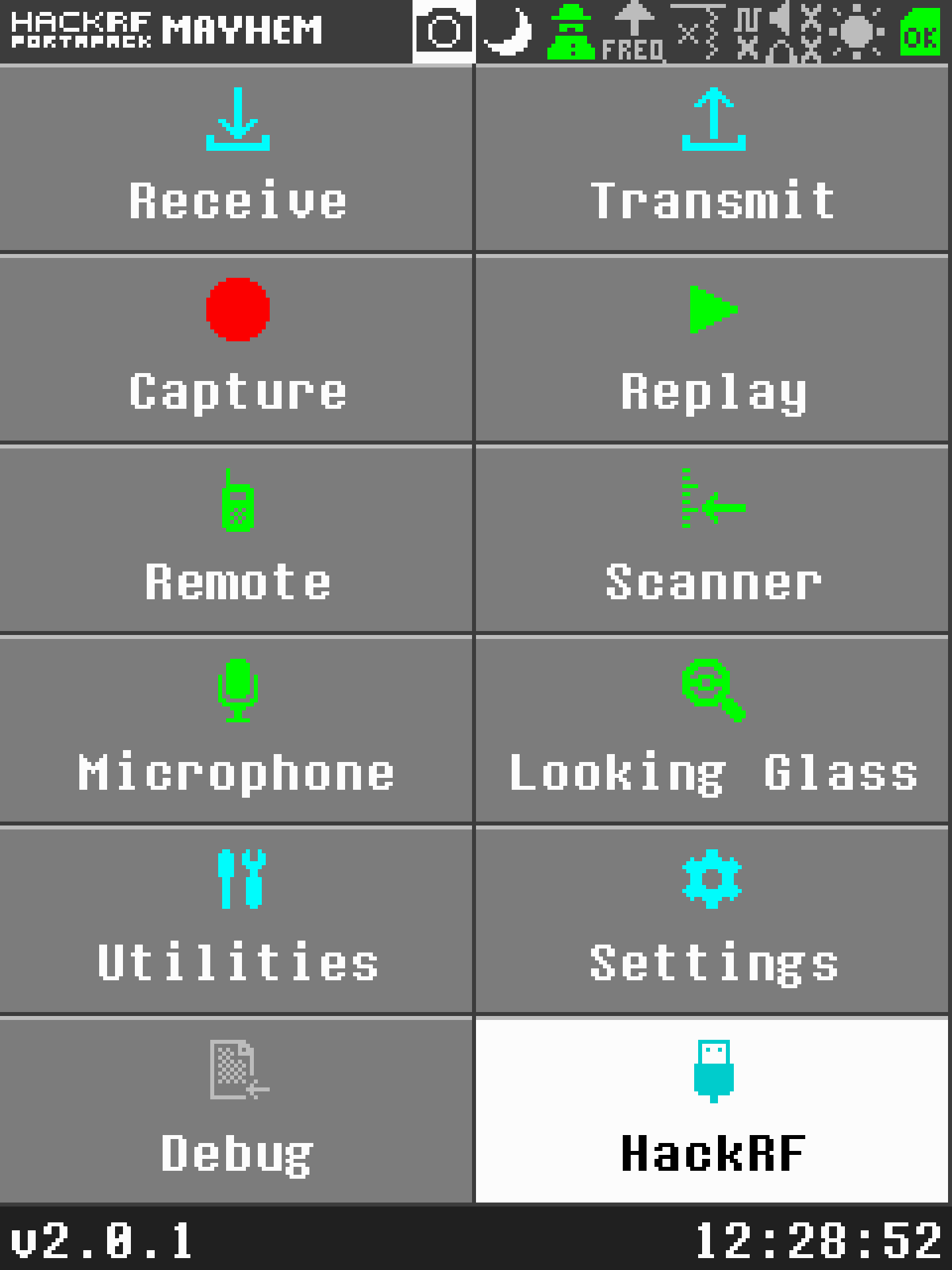}}
\caption{Screenshot \texttt{SCREENSHOTS/SCR\_0000.PNG} saved by the device. The image is enlarged 6 times with nearest-neighbour interpolation.}
\label{fig:screenshot}
\end{figure}

In the example image, all three sources indicate v2.0.1. The file \path{FIRMWARE/portapack-mayhem_v2.0.1.bin} (1,048,576~B) is byte-for-byte identical to the firmware image in the official \texttt{mayhem\_v2.0.1\_\allowbreak FIRMWARE.zip} package and contains the string \texttt{FW version: v2.0.1}. The bottom line of the screenshot \texttt{SCREENSHOTS/SCR\_0000.PNG} reads ``v2.0.1'' (Figure~\ref{fig:screenshot}). The fact that the clock on the screen (12:28:52) matches the FAT creation time of the file to the second also shows that the RTC was running on 2024-10-25. The modification times of the stock SD package files are 2024-04-07 04:24--04:33 UTC, just before the v2.0.1 release time of 04:37:29 UTC. The only exception is \texttt{ADSB/\allowbreak world\_map.bin}, which is dated 2023-05-06.

\FloatBarrier
\subsection{SD card artifact catalogue}\label{sec:catalogue}

Table~\ref{tab:catalogue} lists the user and application artifacts that Mayhem v2.0.1 writes to the SD card, with their naming patterns, formats, the moments at which they are written and their forensic value, each together with its counterpart in the source code. This table is designed as a reference for the examination of this class of device.

{\scriptsize
\setlength{\tabcolsep}{3pt}
\renewcommand{\arraystretch}{0.93}
\begin{xltabular}{\textwidth}{@{}>{\raggedright\arraybackslash}p{0.3cm}P{2.7cm}P{2.3cm}LLP{2.6cm}@{}}
\caption{Mayhem v2.0.1 SD card artifact catalogue.}\label{tab:catalogue}\\
\toprule
\# & Path / pattern & Format & When written & Content and forensic value & Source code~\cite{mayhem} \\
\midrule
\endfirsthead
\multicolumn{6}{@{}l}{\small Table~\ref{tab:catalogue} (continued)}\\
\toprule
\# & Path / pattern & Format & When written & Content and forensic value & Source code~\cite{mayhem} \\
\midrule
\endhead
\bottomrule
\endlastfoot
1 & \texttt{SETTINGS/<app>.ini} & \texttt{key=value} text & Re-created at every exit from the application & That the application was opened at least once; last closing time, last frequency, gain and modulation & \texttt{app\_settings.cpp} (\texttt{SettingsManager} destructor) \\
2 & \texttt{SETTINGS/NAV.INI} & \texttt{autostart\_app=} & When the setting changes & Application started automatically at boot; if set, the device may transmit as soon as it is switched on & \texttt{ui\_navigation.cpp} (\texttt{handle\_\allowbreak autostart}) \\
3 & \texttt{SETTINGS/\allowbreak DATE\_FILEFLAG} & 0-byte flag file & At every boot while the RTC year is 0 & Its creation date is advanced by one day at every boot; ``1980-01-0N'' is a boot counter & \texttt{ui\_navigation.cpp} (\texttt{rtc\_battery\_\allowbreak workaround}) \\
4 & \texttt{CAPTURES/\allowbreak BBD\_NNNN.C16} (or \texttt{.C8}) & Headerless, interleaved int16 I/Q, little-endian & During the capture & Recorded radio signal; duration = size / (4 $\times$ sample rate) & \texttt{ui\_record\_\allowbreak view.cpp}, \texttt{capture\_app.hpp} \\
5 & \texttt{CAPTURES/\allowbreak BBD\_NNNN.TXT} & \texttt{center\_\allowbreak frequency=}, \texttt{sample\_rate=} (position if GPS is available) & At the start of the capture & Centre frequency and sample rate; without it the capture cannot be interpreted & \texttt{metadata\_file.cpp} \\
6 & \texttt{<file>-tmp} (deleted) & Trimmed copy of the source file & While IQ Trim runs & Evidence that a capture was trimmed; the original data may remain in unallocated space & \texttt{iq\_trim.cpp} \\
7 & \texttt{<file>\textasciitilde} & Working copy of the edited file & On the first edit in Notepad; remains after saving & Evidence that text was edited on the device; the \texttt{\textasciitilde} file is \emph{older} than the main file & \texttt{ui\_text\_\allowbreak editor.cpp} \\
8 & \texttt{SCREENSHOTS/\allowbreak SCR\_NNNN.PNG} & 240$\times$320-pixel colour PNG & Via the camera icon in the status bar or the \texttt{screenshot} command in the USB shell & Current screen image: version, clock, open application & \texttt{ui\_navigation.cpp} (\texttt{on\_camera}), \texttt{png\_writer.cpp} \\
9 & \texttt{REMOTES/\allowbreak REMOTE\_NNNN.REM} & Line 1 remote name; following lines \texttt{path,\allowbreak name,\allowbreak icon,\allowbreak background,\allowbreak foreground,\allowbreak fc,\allowbreak fs} & Saving in the Remote application & User-defined buttons and linked IQ captures & \texttt{ui\_remote.cpp} \\
10 & \texttt{PLAYLIST/\allowbreak PLAY\_NNNN.PPL} & \texttt{path[,delay\_ms]} & When a replay list is saved & Captures selected for replay; may contain references to deleted captures & \texttt{ui\_playlist.cpp} \\
11 & \texttt{BLERX/Lists/\allowbreak <name>.CSV} & \texttt{MAC ADV\_DATA 50} (no header) or list export with header & ``Save'' in the packet details, or list export & Captured BLE advertising packets; the file name is given by the user & \texttt{ble\_rx\_app.cpp} \\
12 & \texttt{BLETX/*.TXT} & \texttt{MAC ADV\_DATA COUNT} & When a file is created in the BLE transmit application & BLE packets prepared for transmission & \texttt{ble\_tx\_app.cpp} \\
13 & \texttt{FIRMWARE/*.bin} & Firmware image & Copied by the user & Indirect evidence of the installed version; hash comparison with the official release & --- \\
14 & \texttt{APPS/*.ppma} & External application & With the SD package & Application set matching the version; verified against the official package & \texttt{ui\_external\_\allowbreak items\_menu\_\allowbreak loader.cpp} \\
15 & \texttt{ADSB/\allowbreak world\_map.bin} & 4~B header (width, height) + 16-bit colour pixels & With the SD package & 2,147,483,652~B = 4 + 32,768 $\times$ 32,768 $\times$ 2; a different size indicates corruption & \texttt{generate\_world\_\allowbreak map.bin.py}, \texttt{ui\_geomap.cpp} \\
16 & \texttt{System Volume Information/} & Windows directory & When the card is inserted into Windows & Windows connection; volume identifier in the \texttt{IndexerVolumeGuid} file & --- \\
17 & \texttt{.Trash-<uid>/} & freedesktop trash can~\cite{trashspec} & When deleted on a Linux desktop & Name of the deleted file, deletion time in local time (\texttt{DeletionDate}) and user identifier (uid) & --- \\
\end{xltabular}
}

Among the artifacts in the catalogue, the settings files are the richest source of information on application use. Each settings file belongs to a specific application (Table~\ref{tab:settings}). The existence of the file shows that the application was opened and closed at least once. Because Mayhem re-creates the file at every exit, its creation and modification times are renewed together at every write. The file time therefore shows the last closure of the application, not its first use; for example, \texttt{rx\_capture.ini} is dated 2024-12-19, whereas the Capture application made captures BBD\_0003--BBD\_0011 on 2024-11-30. Values that carry notable user input show that the application was configured. A settings file proves only that the application was opened; it does not prove that a transmission was made. The common keys are \texttt{rx\_frequency}/\texttt{tx\_frequency} in Hz, \texttt{lna} and \texttt{vga} for the receiver gain in dB, \texttt{rx\_amp}/\texttt{tx\_amp} for the additional signal amplifier, \texttt{tx\_gain} in dB, \texttt{sampling\_rate} and \texttt{baseband\_bandwidth} in Hz, and \texttt{modulation} for the modulation type. The modulation values are 0 for amplitude modulation (AM), 1 for narrowband frequency modulation (NFM), 2 for wideband frequency modulation (WFM), 3 for spectrum and 4 for capture.

Of the 28 settings files in the example image, 24 were last written between 20:57 and 21:53 on 2024-12-16, mostly a few minutes apart (0.1--10.3~min; median 1.3~min). This indicates that the applications in the menu were opened and closed one after another in that session. The values were reviewed manually. Seven files carry notable user input, namely a specific frequency, list name, message text or raised gain. These files are given in Table~\ref{tab:settings}. For example, \texttt{tx\_pocsag.ini} is the only file in which the additional signal amplifier was switched on and the gain was raised.

\begin{table}[!htb]
\centering
\caption{Settings files with notable user input in the example image; last write time by the device clock.}
\label{tab:settings}
\small
\begin{tabularx}{\textwidth}{@{}P{2.8cm}P{2.9cm}LP{2.4cm}@{}}
\toprule
File & Application (menu) & Last value & Time \\
\midrule
\texttt{rx\_capture.ini} & Capture & 433.920~MHz; 4~MS/s & 2024-12-19 22:57 \\
\texttt{tx\_replay.ini} & Replay & 433.920~MHz; gain 35~dB & 2024-12-19 22:58 \\
\texttt{tx\_remote.ini} & Remote & 269.84375~MHz; \texttt{remote\_path=\allowbreak REMOTES/\allowbreak REMOTE\_0000.REM} & 2024-12-16 21:33 \\
\texttt{rx\_scanner.ini} & Scanner & 476.8125~MHz; list \texttt{CB\_UHF\_AUSTRALIA} & 2024-12-16 21:35 \\
\texttt{RX\_GLASS.INI} & Looking Glass & 5.16--5.90~GHz sweep & 2024-12-16 21:36 \\
\texttt{TX\_MORSE.INI} & Morse & 2.402~GHz; \texttt{message=root0ne} & 2024-12-16 21:03 \\
\texttt{tx\_pocsag.ini} & POCSAG TX (Post Office Code Standardisation Advisory Group pager transmitter) & 466.175~MHz; \texttt{tx\_amp=1}, \texttt{tx\_gain=40} & 2024-12-16 21:13 \\
Other 21 files & Audio, BLE Rx, BLE Tx, GPS Sim, ADS-B TX, Microphone, Notepad, Settings and 13 other applications & No notable user input & 2024-12-16 20:57 -- 2024-12-20 09:33 \\
\bottomrule
\end{tabularx}
\end{table}

Among the other user artifacts, the file \texttt{REMOTES/REMOTE\_0000.REM} contains a remote with the unchanged name \texttt{<Unnamed Remote>} and two buttons: one keeps its default name \texttt{<EMPTY>}, and the device user named the other ``duzgun''. The capture path of both buttons is empty, so this remote cannot trigger a transmission. \texttt{BLERX/Lists/HII.CSV} is a single advertising packet saved with the ``Save'' function of the BLE Rx application. Its media access control (MAC) address is \texttt{0E:98:19:C6:\textbullet\textbullet:\textbullet\textbullet}. The packet is a Connected Devices Platform beacon that carries the Microsoft company identifier and that Windows computers broadcast to discover nearby devices~\cite{mscdp}, \cite{btassigned}. The device type field (9) indicates a Windows~10 desktop. This points to a Windows computer having been nearby at the time of capture. \texttt{GPS/gps-sdr-sim} is an x86-64 Linux build of the osqzss/gps-sdr-sim tool~\cite{gpssdrsim} compiled on Ubuntu with the GNU Compiler Collection (GCC) 13.2.0. This file does not run on the device. Its presence on the card shows that the GPS signal-generation tool was compiled on a computer and copied to the card. In an examination, such files help to identify the operating system and the software tools of the computer used together with the device.

\FloatBarrier
\subsection{Interpreting timestamps and the timeline}\label{sec:time}

FAT32 timestamps carry no time-zone information and have a resolution of 2 seconds~\cite{fat32spec}. Each timestamp must therefore be read according to the rule of the system that wrote it (Table~\ref{tab:timestamps}). The device takes its time from the RTC of the LPC4320. If the RTC backup battery (K1) is not fitted or is flat, the year is read as 0 at boot. In this case, Mayhem advances the date of the \texttt{SETTINGS/DATE\_FILEFLAG} file by one day at every boot and sets the clock to this date. ``1980-01-0N'' dates are therefore not calendar days but the number of boots without a battery. The real date of this period can be bounded by relative evidence such as the file numbers and the cluster allocation order, that is, the order in which space was allocated to files on the card. The FatFs library used by Mayhem and the Linux \texttt{vfat} driver allocate space for a new file immediately after the most recently used location. Cluster numbers therefore roughly indicate the order in which the files were written.

\begin{table}[!htb]
\centering
\caption{Timestamp writing rules and time-zone inference in the example image.}
\label{tab:timestamps}
\footnotesize
\begin{tabularx}{\textwidth}{@{}P{2.6cm}P{3.0cm}LL@{}}
\toprule
Writer & Rule & Evidence in the example image & Inference \\
\midrule
Mayhem (device) & RTC time; without a battery, the \texttt{DATE\_FILEFLAG} counter & Last device write \texttt{NAV.INI} 09:33:42; image completed at 08:42:28 UTC & Device clock is at least 51~min ahead of UTC, so it is not UTC; consistent with Türkiye local time (UTC+3) \\
Windows & Local time & \texttt{System Volume Information} 2024-08-23 22:51 & Local time \\
Linux \texttt{vfat} (this computer)~\cite{vfatdoc} & UTC (verified with the trash record) & \texttt{.Trash-1000} FAT 21:55:36; \texttt{trashinfo} \texttt{DeletionDate} 00:55:36 (local) & Linux computer at UTC+3 \\
Acquisition computer & ZIP archive's local-time and UTC fields & 11:42:28 local = 08:42:28 UTC & Computer at UTC+3; acquisition completed at 2024-12-20 08:42:28 UTC \\
Official SD package & UTC in the release pipeline & Stock files 04:24--04:33 & Gives no information about the user \\
\bottomrule
\end{tabularx}
\end{table}

\begin{figure}[!htb]
\centering
\figfile{\includegraphics[width=\textwidth]{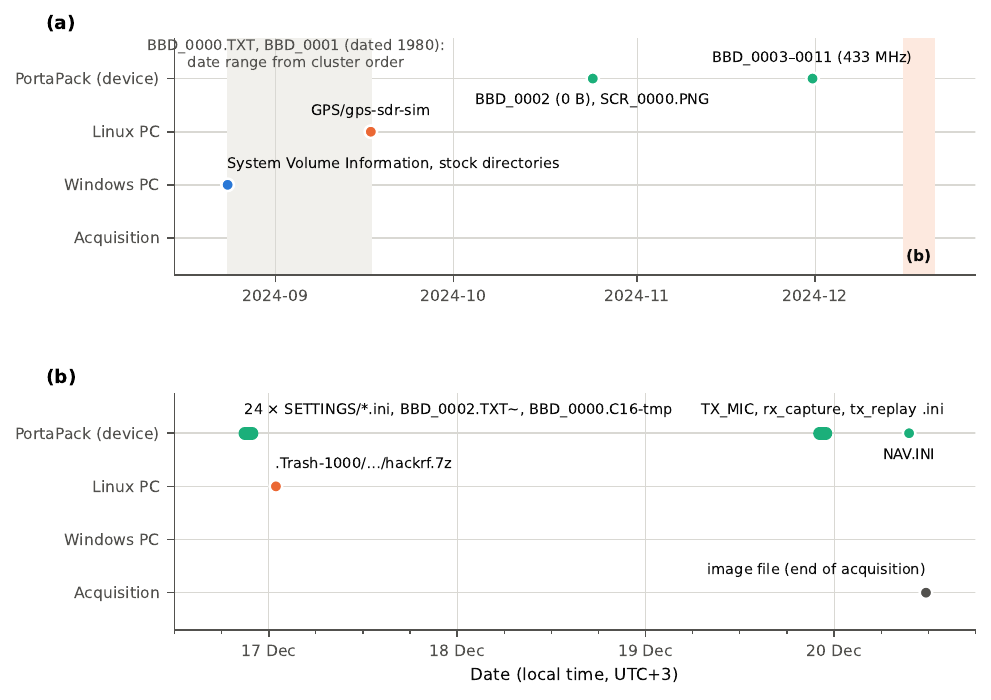}}
\caption{Timeline of the example image (local time, UTC+3): (a) August--December 2024, (b) 16--20 December 2024.}
\label{fig:timeline}
\end{figure}

In the example image, the \texttt{CAPTURES/} directory, \texttt{BBD\_0000.TXT} and the BBD\_0001 capture are dated 1980-01-01, \texttt{LOGS/} 1980-01-02 and \texttt{DATE\_FILEFLAG} 1980-01-05. \texttt{BBD\_0000.C16} carries a later date because IQ Trim rewrote it on 2024-12-16. Accordingly, the device was switched on about five times without a battery. The real date of these captures was bounded by two pieces of relative evidence. First, the BBD\_0001 capture and the \texttt{LOGS/} directory occupy the card up to cluster 79,522. In the very next cluster, 79,523, a computer wrote the \texttt{gps-sdr-sim} file on 2024-09-17; these captures were therefore made before that date. Second, BBD\_0000 and BBD\_0001 were written into the space freed by the SD package ZIP file that had been copied to the card on 2024-08-23 and later deleted; these captures were therefore made after the ZIP was deleted. This inference assumes that the ``next free cluster'' hint kept in the FSInfo sector of the card is also used across different systems when space is allocated to new files. Because the screenshot shows that the RTC was running on 2024-10-25, the battery-less period ended before that date. Accordingly, the 1980-dated captures were most probably made between 2024-08-23 and 2024-09-17. Figure~\ref{fig:timeline} shows the events in local time. Each writing system has its own row: device, Windows, Linux and acquisition. The system to which each timestamp belongs was determined according to the rules in Table~\ref{tab:timestamps}, and the timestamps written by Linux in UTC were converted to local time. In the figure, each dot represents a timestamp and each bar the interval covered by consecutive timestamps. The grey region is the date range computed from the cluster order for the 1980-dated captures.

\FloatBarrier
\subsection{Detecting and recovering deleted, trimmed and edited data}\label{sec:deleted}

In Mayhem, a capture can change in three ways: deletion, trimming with IQ Trim and editing with Notepad. Each of these leaves a distinctive trace:
\begin{enumerate}
  \item When a file is deleted, FAT32 sets the first byte of the file's directory entry to \texttt{0xE5} and marks the file's clusters as free; the data themselves remain in place until new data are written over them~\cite{carrier2005}, \cite{fat32spec}. A C16 capture is a capture file consisting of 16-bit IQ samples. Because Mayhem writes captures to large, contiguous runs of clusters, the data of a deleted C16 capture can be recovered contiguously from the free cluster range between neighbouring captures. Playlists (\texttt{*.PPL}) and remote files (\texttt{*.REM}) preserve the names of deleted captures.
  \item When IQ Trim is used, the trimmed data are written to a \texttt{<name>-tmp} file, the original is deleted and the \texttt{-tmp} file is renamed. The deleted \texttt{-tmp} entry is evidence of trimming. Because the clusters of the original capture are freed, the trimmed-away part can be recovered unless it has been overwritten. To find it, a portion of the live, undeleted file is searched for in unallocated space; the start offset of the match gives the length of the discarded part.
  \item On the first edit in Notepad, a \texttt{<name>\textasciitilde} copy is created and the changes are written to this copy. On saving, the main file is deleted and re-created from the copy. The \texttt{\textasciitilde} file is therefore older than the main file. This ordering shows that the editing was done on the device and allows it to be distinguished from the traces of computer editors.
\end{enumerate}

Table~\ref{tab:deleted} summarizes the main deleted and modified entries in the example image. Three results stand out.

\begin{table}[!htb]
\centering
\caption{Deleted and modified entries.}
\label{tab:deleted}
\footnotesize
\begin{tabularx}{\textwidth}{@{}P{3.6cm}P{2.5cm}LL@{}}
\toprule
Entry & Time & Status & Evidential value \\
\midrule
\texttt{CAPTURES/BBD\_0009.C16} (no directory entry) & 2024-11-30 $\approx$15:30--15:37 & Last 2,000 clusters ($\approx$4.1~s) recovered; the head may be overwritten & An additional key press of the same transmitter \\
\texttt{CAPTURES/\allowbreak BBD\_0000.C16-tmp} & 2024-12-16 21:43:40 & Deleted; same clusters and content as the live file & Use of IQ Trim \\
Original BBD\_0000 data & 1980 (battery-less period) & 258,608~B recovered, 16,384~B overwritten & The discarded part contains only noise and a continuous-wave (CW) carrier \\
1st Trim output (92,439--92,687) & 2024-12-16 21:42:54--21:43:40 & Deleted; identical to the live file & Trim was run twice \\
\texttt{CAPTURES/BBD\_0002.C16} & 2024-10-23 10:56:36 & 0 bytes, deleted & A capture that was started and ended without data being written \\
\texttt{CAPTURES/BBD\_0006.TXT} & 2024-11-30 15:28:36 & Deleted; its cluster reused by \texttt{NOTEPAD.INI} & Metadata of the capture lost; frequency unknown \\
\texttt{BLETX/\allowbreak TXTdataFileTemp.} & 2024-12-16 21:00:04 & Deleted; its cluster reused by \texttt{TX\_MORSE.INI} & BLE TX temporary file \\
\texttt{/hackrf.7z} and \texttt{.Trash-1000/\allowbreak files/\allowbreak hackrf.7z} & 2024-12-17 00:55 (local) & 0 bytes; no 7z signature in the image & Attempt to archive the card on Linux \\
Mayhem SD package ZIP remnant & 2024-08-23 & 516,700,350~B; 173 entries (17 directories, 156 files); 155 of the 156 files pass the CRC check & Same size as the official \texttt{COPY\_TO\_SDCARD.zip}; the 155 files that pass the CRC are byte-for-byte identical to the stock copies on the card \\
Orphan directory cluster 917,509 & 2024-08-23 23:13 & 15 deleted subdirectory entries & The ZIP was extracted into a subfolder and its contents moved to the root \\
\bottomrule
\end{tabularx}
\end{table}

\begin{figure}[!htb]
\centering
\figfile{\includegraphics[width=\textwidth]{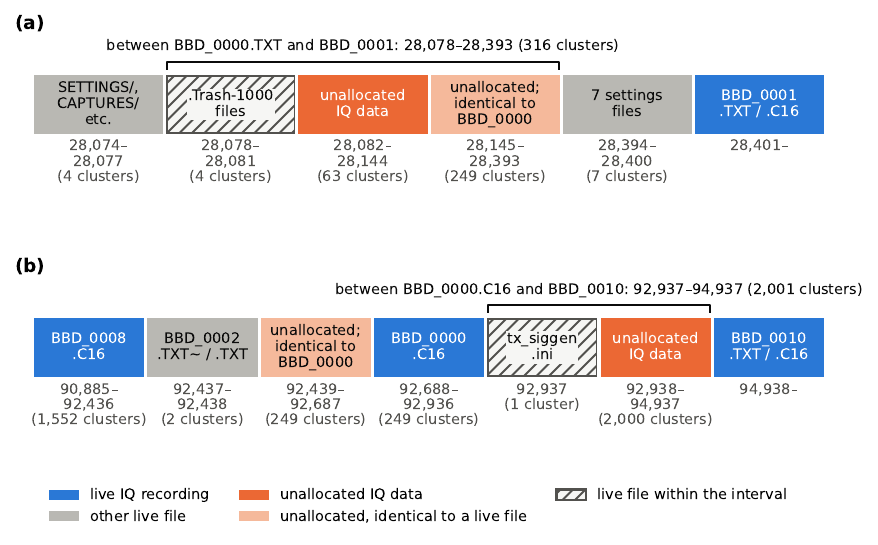}}
\caption{Content of two cluster ranges: (a) 28,078--28,393, (b) 92,937--94,937. Block widths are not to scale; cluster numbers are given below the blocks.}
\label{fig:clustermap}
\end{figure}

First, the final part of a deleted capture was recovered. The file \texttt{PLAYLIST/\allowbreak PLAY\_0000.PPL} refers to the capture \texttt{/CAPTURES/\allowbreak BBD\_0009.C16}, which is not present in the directory. The range from cluster 92,436, where BBD\_0008 ends, to cluster 94,938, where BBD\_0010 begins, was partly reused on 2024-12-16: the BBD\_0002 metadata files, the first Trim output, \texttt{BBD\_0000.C16} and \texttt{tx\_siggen.ini} were written to the clusters at its start. The remaining clusters 92,938--94,937 (2,000 clusters, $\approx$4.1~s) contain valid IQ data (Figure~\ref{fig:clustermap}(b)). Because no directory entry remains, it cannot be verified conclusively which capture these data belong to; however, their location and the missing capture in the playlist indicate that they are very probably the final part of BBD\_0009. About 1~s at the beginning of the capture may have been overwritten. In the recovered part, a key press of the same transmitter as in the other 433~MHz captures was found (Section~\ref{sec:iq}).

\begin{figure}[!htb]
\centering
\figfile{\includegraphics[width=\textwidth]{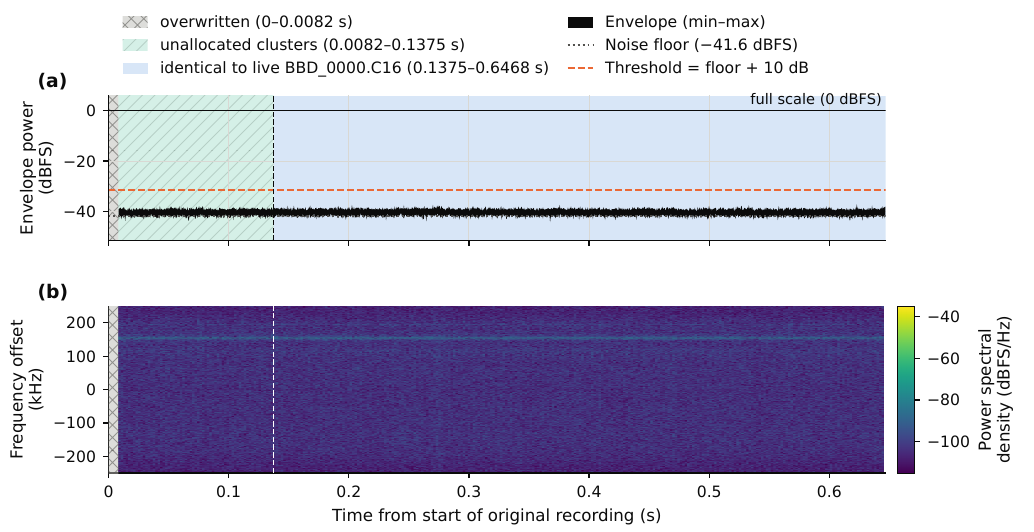}}
\caption{IQ data read from clusters 28,078--28,393: (a) envelope power, (b) spectrogram.}
\label{fig:bbd0000}
\end{figure}

Second, the data discarded with IQ Trim were recovered. The current content of the file BBD\_0000.C16 was found byte-for-byte in unallocated space; however, this match begins at byte 274,992 of the original capture. Accordingly, the 0.1375~s part at the beginning of the 1980-dated original capture was discarded with IQ Trim. The first 0.0082~s of this part was later overwritten by Linux trash files, and the rest was recovered (Figure~\ref{fig:clustermap}(a), Figure~\ref{fig:bbd0000}). The recovered part contains only noise and a CW carrier, a constant signal that carries no information. The cluster allocation order shows that trimming was run twice on 2024-12-16 (Table~\ref{tab:deleted}).

Figure~\ref{fig:clustermap} shows the content of these two cluster ranges. In the figure, ``unallocated IQ data'' refers to clusters that are marked as free in the file system but contain IQ data, and ``identical to a live file'' refers to clusters that are byte-for-byte identical to an undeleted file.

Figure~\ref{fig:bbd0000} shows the power and the frequency content of the data read from these clusters over time. The centre frequency (fc) of the capture is 269.84375~MHz and the sample rate (fs) is 500~kS/s. The figure shows the overwritten part in the first 0.0082~s, the recovered part between 0.0082 and 0.1375~s, and the part identical to the live BBD\_0000.C16 file between 0.1375 and 0.6468~s.

Third, it was determined that text had been edited on the device. \texttt{BBD\_0002.TXT\textasciitilde} was written 14~s before the main file, and the contents of the two files are identical; free text has been added to the metadata file. The fact that the \texttt{\textasciitilde} file is older and that the \texttt{NOTEPAD.INI} file was written 20~s later shows that the editing was done with Notepad on the device. Because the edit corrupted the line endings, the metadata file has become unreadable.

\FloatBarrier
\subsection{Characterizing and decoding IQ captures}\label{sec:iq}

The examination of a C16 capture consists of four steps:
\begin{enumerate}
  \item The centre frequency (fc) and the sample rate (fs) are read from the information file next to the capture; the duration of the capture is calculated from the file size and compared with the file times.
  \item The envelope, the PSD and the spectrogram are used to determine whether the capture contains a signal, how wide a frequency band the signal occupies and how far it lies from the centre frequency.
  \item If the envelope shows only two levels, on and off, and consists of pulses, the signal is classified as OOK. The pulse and gap durations are then grouped, and the line code, that is, the rule by which the bits are converted into pulses and gaps, is determined.
  \item The bit sequence is extracted, and the result is compared with an independent decoder such as \texttt{rtl\_433}.
\end{enumerate}

The line code is identified by looking at which lengths of pulses and gaps occur next to each other. In pulse-width modulation (PWM), each bit is a single pulse--gap pair, and only two combinations are seen: short pulse--long gap and long pulse--short gap. In Manchester coding, by contrast, both short and long pulses can be combined with both short and long gaps, so four combinations are seen. This distinction matters, because if the line code is assumed wrongly, the number and the values of the decoded bits come out wrong. ADC saturation, that is, the signal being clipped at the highest value the ADC can measure, indicates that the transmitter was very close to the receiver. Here, the saturated envelope reaches about +3~dBFS. Captures that belong to the same transmitter are recognized by identical pulse durations and an identical fixed part at the beginning of the frames. If the frame has a part that changes with every press, the transmitter uses a rolling code; that is, it sends a different code with every press and is thereby protected against being recorded and replayed.

The results for the ten live captures in the example image are summarized in Table~\ref{tab:iq}. For each capture, the table gives the centre frequency, the duration, the peak power, the signal class and the number of presses found.

\begin{table}[!htb]
\centering
\caption{Summary of the IQ captures in the example image (fs = 500~kS/s).}
\label{tab:iq}
\footnotesize
\begin{tabularx}{\textwidth}{@{}P{2.3cm}P{1.8cm}P{1.4cm}P{1.3cm}P{2.6cm}L@{}}
\toprule
Capture & fc (MHz) & Duration (s) & Peak (dBFS) & Class & Finding \\
\midrule
BBD\_0000 & 269.844 & 0.51 & $-37$ & No signal & Only an unmodulated CW carrier at +153.9~kHz \\
BBD\_0001 & 269.844 & 104.7 & $-29$ & No data & Electrical interference and two narrow CW lines; no decodable data \\
BBD\_0003 & 433.920 & 2.97 & +2.8 & OOK, Manchester & 1 press (key 2) \\
BBD\_0004 & 433.920 & 9.68 & +2.8 & OOK, Manchester & 2 presses (key 2); also weak, undecodable pulses \\
BBD\_0005 & 433.920 & 1.03 & $-43$ & Noise & --- \\
BBD\_0006 & (metadata lost) & 3.46 & $-42$ & Noise & --- \\
BBD\_0007 & 433.000 & 5.96 & $-21.5$ & OOK, Manchester (weak) & Same transmitter (key 4), through aliasing \\
BBD\_0008 & 434.000 & 3.18 & $-42$ & Noise & --- \\
BBD\_0009 (recovered) & unknown ($\approx$434.0) & 4.10 & +2.0 & OOK, Manchester & 1 press (key 4) \\
BBD\_0010 & 433.900 & 2.51 & +2.5 & OOK, Manchester & 1 press (key 4) \\
BBD\_0011 & 433.920 & 11.08 & +2.8 & OOK, Manchester & 4 presses (key 4) \\
\bottomrule
\end{tabularx}
\end{table}

\begin{figure}[!htb]
\centering
\figfile{\includegraphics[width=\textwidth]{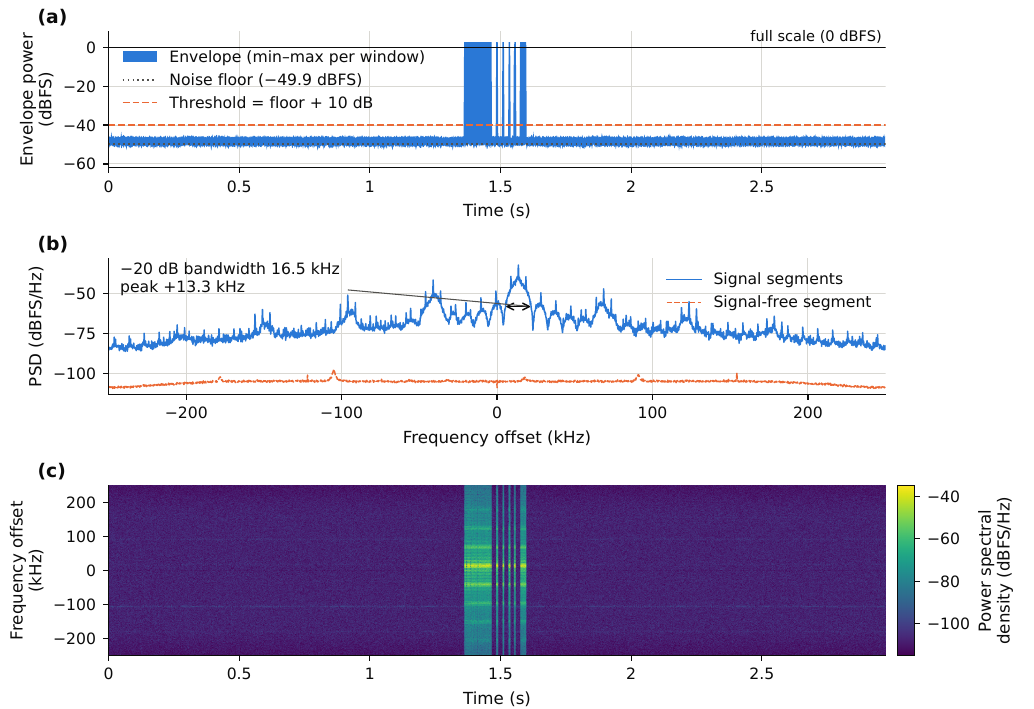}}
\caption{Capture BBD\_0003 (fc = 433.920~MHz, fs = 500~kS/s): (a) envelope power, (b) power spectrum of the signal and noise regions, (c) spectrogram.}
\label{fig:overview}
\end{figure}

The two captures at 269.84~MHz contain no decodable signal. These captures were also tried with decoders for the POCSAG and FLEX pager protocols and for audio frequency-shift keying (AFSK), but no data were obtained. Four of the 433--434~MHz captures and the recovered BBD\_0009 data contain OOK pulse packets that drive the ADC into saturation. Figure~\ref{fig:overview} shows the envelope, the power spectrum and the spectrogram of one of these captures, BBD\_0003. The carrier lies 13.3~kHz above the centre frequency of the capture ($\approx$433.933~MHz), and the signal occupies a band about 16.5~kHz wide when measured 20~dB below its highest power. The saturation also creates additional peaks, called intermodulation products, on both sides of the carrier in the power spectrum (Figure~\ref{fig:overview}(b)).

\begin{figure}[!htb]
\centering
\figfile{\includegraphics[width=\textwidth]{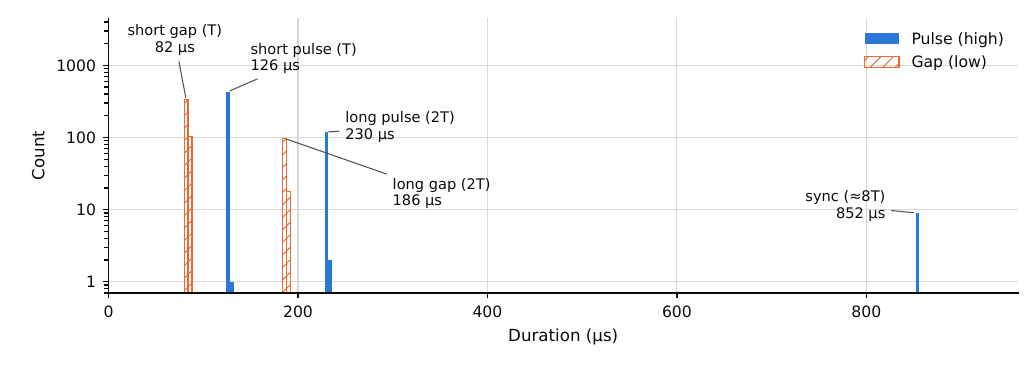}}
\caption{Distribution of pulse and gap durations in BBD\_0003. The vertical axis is logarithmic. The labels show the median duration of each group.}
\label{fig:durations}
\end{figure}

To determine the line code, the durations of all pulses and gaps in BBD\_0003 were measured (Figure~\ref{fig:durations}). Pulses and gaps occur with two different lengths. With a base duration of $T = 104$~$\mu$s, these lengths are approximately $T$ and $2T$: the pulses last $\approx$126 and 230~$\mu$s, and the gaps $\approx$82 and 186~$\mu$s. The bit rate is 4,808~bit/s. The observation of all four pulse--gap combinations shows that the line code is Manchester. The automatic pulse analyser of \texttt{rtl\_433} guessed the signal to be PWM, but the Manchester decoder of the same tool, except for the weak BBD\_0007 capture, gave the same bits after the 16-bit header as the decoding of this study (Table~\ref{tab:validation}).

\begin{figure}[!htb]
\centering
\figfile{\includegraphics[width=\textwidth]{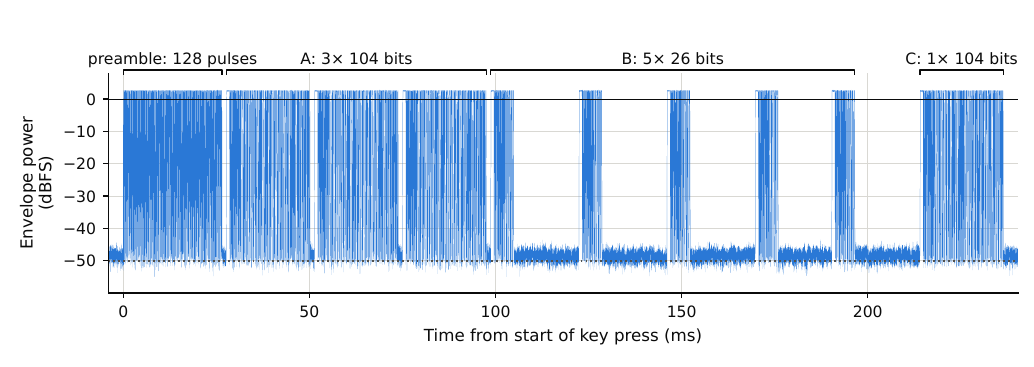}}
\caption{Envelope power and frame groups of one key press in BBD\_0003.}
\label{fig:packet}
\end{figure}

Figure~\ref{fig:packet} shows the frame structure of one key press in the same capture. Each pulse packet consists of a 128-pulse start sequence called the preamble, three identical 104-bit A frames, 26-bit B frames repeated at intervals of $\approx$23.7~ms and one 104-bit C frame. This press contains 5 B frames, and their number increases with the duration of the packet. This structure indicates that A is the data frame sent when the key is pressed, B is the repeat frame sent while the key is held down, and C is the frame sent when the key is released. One packet therefore corresponds to one key press.

\begin{figure}[!htb]
\centering
\figfile{\includegraphics[width=\textwidth]{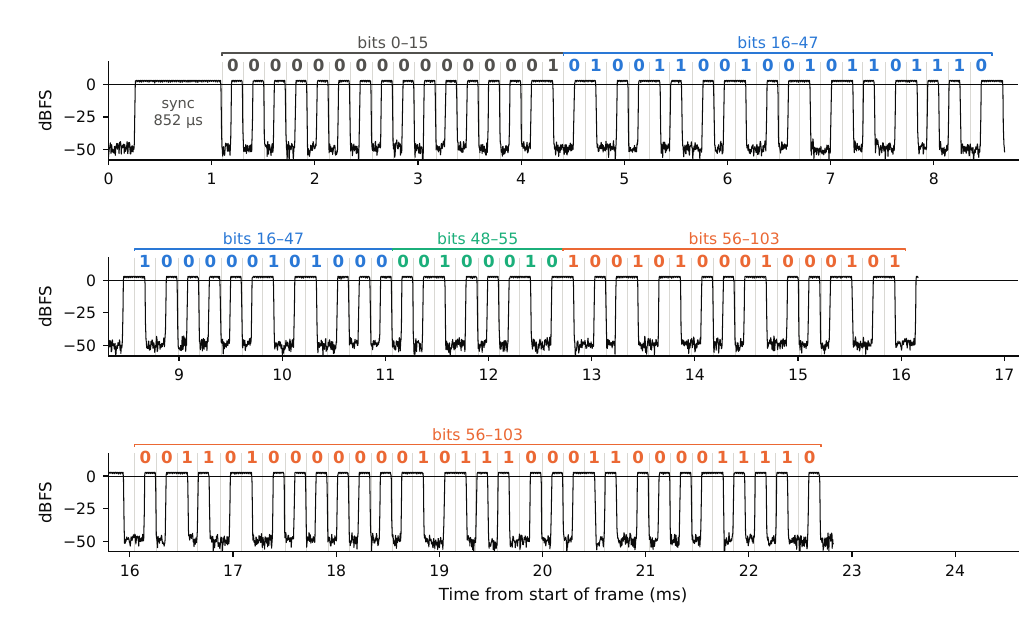}}
\caption{Manchester decoding of the first A frame in BBD\_0003. The frame is split into three rows; the colours indicate the bit ranges (0--15, 16--47, 48--55, 56--103).}
\label{fig:bits}
\end{figure}

Figure~\ref{fig:bits} shows the bit-by-bit decoding of the first A frame. In Manchester coding, each bit is read from the direction of the transition in its middle: a high-to-low transition is 1 and a low-to-high transition is 0. In this frame, $T = 104$~$\mu$s and the bit rate is 4,808~bit/s. The fields of the data frame are a 16-bit header, a 32-bit fixed block, a 4-bit status field, a 4-bit key field and a 48-bit variable block.

\begin{figure}[!htb]
\centering
\figfile{\includegraphics[width=\textwidth]{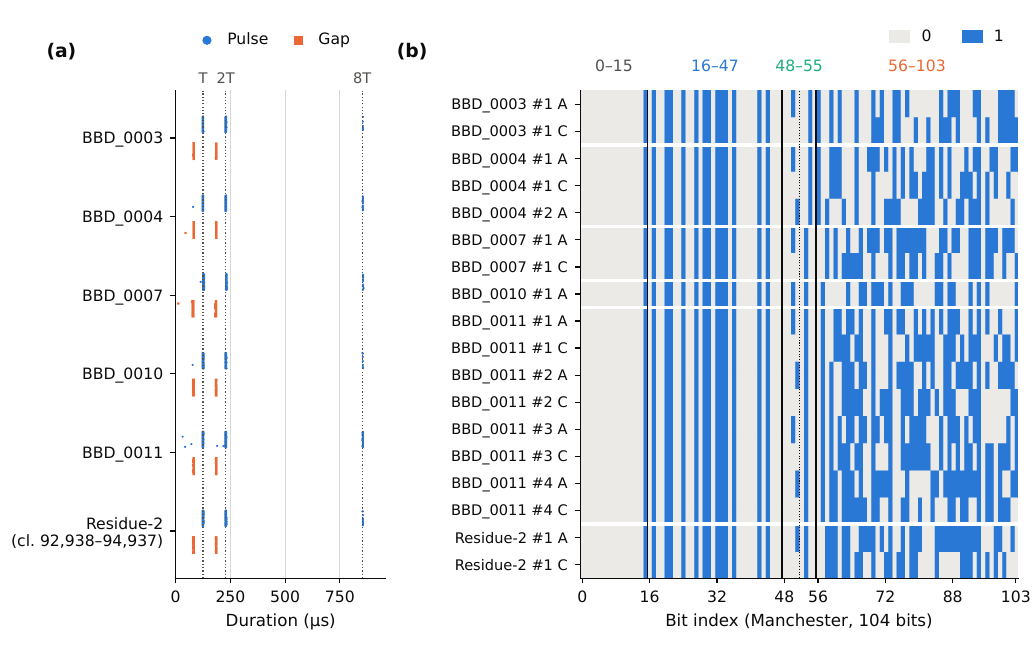}}
\caption{Comparison of the six sources: (a) pulse and gap durations per source, with dashed lines at $T$, $2T$ and $8T$, (b) bits of the A and C frames of each press. Vertical lines in (b) are field boundaries.}
\label{fig:single}
\end{figure}

Six sources were tested to see whether they belong to the same transmitter: BBD\_0003, BBD\_0004, BBD\_0007, BBD\_0010, BBD\_0011 and the recovered BBD\_0009 data, labelled ``Residue-2'' in the figure. The pulse durations and the frame bits of all sources were compared (Figure~\ref{fig:single}). The pulse and gap durations fall into the same groups in all six sources (Figure~\ref{fig:single}(a)). Bits 16--47 are identical in all 37 complete frames of the six sources, whereas bits 56--103 differ with every press (Figure~\ref{fig:single}(b)). The unchanging 32-bit block is consistent with a device identifier, and the 48-bit block that changes with every press is consistent with a rolling code. The status field alternates between 2 and 1 on successive presses and becomes 0 on release. The key field is 2 in BBD\_0003 and BBD\_0004 and 4 in the other captures; that is, two different keys of the same remote were captured.

This structure is consistent with a rolling-code car or garage remote. It is not consistent with fixed-code remotes that send the same code on every press, such as remotes with PT2262/EV1527 chips. None of the protocols registered in \texttt{rtl\_433} recognized this signal; the brand of the transmitter could therefore not be determined. BBD\_0007 (433.000~MHz) contains a copy of the same transmitter that is about 24~dB weaker. Because this capture was tuned to 433.000~MHz, the transmitter signal at 433.933~MHz falls outside the frequency range of the capture. The signal nevertheless appears at $-68.5$~kHz as an attenuated reflection, because the filter applied before capture does not suppress this frequency sufficiently. This effect is called aliasing; the computed value is $-66.7$~kHz. Because the information file of the recovered BBD\_0009 data is lost, their centre frequency is unknown; the fact that the signal reaches saturation, unlike the weak aliased copy, suggests that the capture was very probably tuned to 434.000~MHz, like BBD\_0008.

\FloatBarrier
\section{Validation}\label{sec:validation}

The findings were validated with independent methods. Table~\ref{tab:validation} lists each main claim, the method that tests it and the result.

\begin{table}[!htb]
\centering
\caption{Validation methods and results.}
\label{tab:validation}
\footnotesize
\begin{tabularx}{\textwidth}{@{}P{3.2cm}LL@{}}
\toprule
Claim & Method & Result \\
\midrule
Original evidence unchanged & hashdeep (MD5, SHA-256) at the start and end of the examination & Matched \\
File list correct & Comparison of Sleuth Kit \texttt{fls} with the independent FAT32 parser & 216 files, 31 directories and 6 deleted file records in the directory tree; consistent \\
Stock content not modified & Byte-for-byte comparison of the 156 files in the embedded ZIP remnant with their copies on the card & 155 identical; the only different file is \texttt{ADSB/\allowbreak world\_map.bin}, whose copy in the ZIP remnant fails the CRC check \\
Firmware v2.0.1 & \texttt{cmp} against the official release package; screenshot; release time & Three sources consistent \\
Artifact interpretations & Mapping of every artifact to the Mayhem v2.0.1 source code (Table~\ref{tab:catalogue}) & All 17 artifacts consistent with the source \\
Captures made in real time & Difference between the duration computed from the size and the creation and modification times & At most 3.3~s for all captures except BBD\_0000, which was rewritten by IQ Trim (5.5~s); FAT time resolution 2~s \\
Notepad and Trim interpretation & File operation order in the source code versus timestamps and cluster order & Consistent \\
Demodulator agrees with \texttt{rtl\_433} & Comparison with the \texttt{rtl\_433} decoding on the Mayhem SD package sample captures (\texttt{SAMPLES/}): 17 Aurora RGB captures (PT2262 type, 25 bits) and a Tesla charge port & 17/17 same house code and command; Tesla identical except for the first bit (Figure~\ref{fig:validation}) \\
Manchester decoding of the 433~MHz captures & Comparison with the \texttt{rtl\_433} Manchester decoder (\texttt{OOK\_MC\_ZEROBIT}, 104~$\mu$s) & In 154 of the 163 frames (data, repeat and release frames) of this study's decoding, all bits after the 16-bit header are identical; the remaining 9 frames are in the weak BBD\_0007 capture, in which \texttt{rtl\_433} decoded no complete frame \\
User captures are not copies of stock samples & SHA-256 and envelope cross-correlation & $|r| \le 0.21$ (0.48--0.81 among themselves) \\
\bottomrule
\end{tabularx}
\end{table}

\begin{figure}[!htb]
\centering
\figfile{\includegraphics[width=\textwidth]{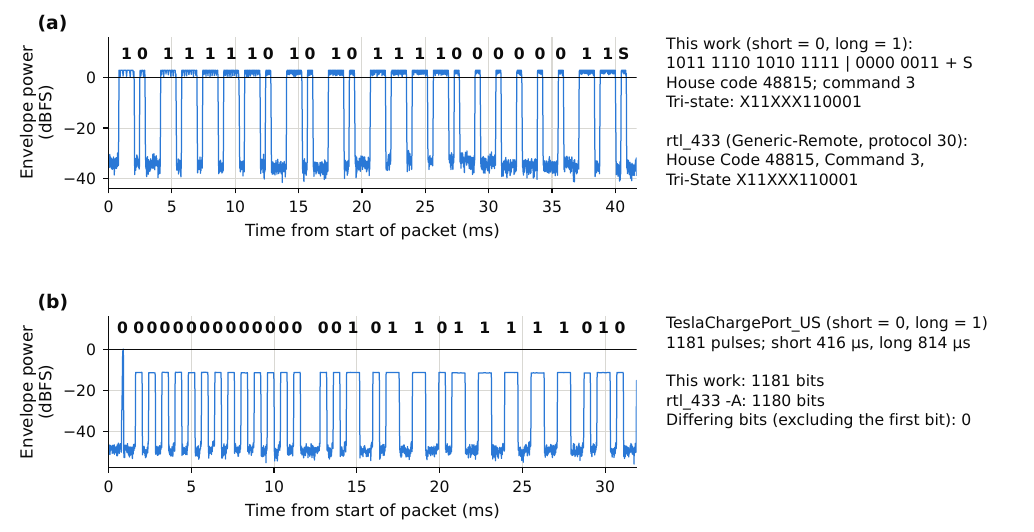}}
\caption{Comparison of the decoding method with \texttt{rtl\_433} on Mayhem sample captures: (a) Aurora RGB remote POWER capture, (b) Tesla charge port capture.}
\label{fig:validation}
\end{figure}

The demodulation method was additionally tested on the \texttt{SAMPLES/} captures of the Mayhem SD package by comparison with \texttt{rtl\_433}, an independent decoder. Figure~\ref{fig:validation} compares the decoding of this study with the \texttt{rtl\_433} output for two sample captures. In the Aurora RGB remote sample capture (433.920~MHz), the 25-bit code, read with short pulses as 0 and long pulses as 1, is identical to the \texttt{rtl\_433} output. In the Tesla charge port sample capture (315.000~MHz), the bit sequences of the two methods differ only in the first bit (1,181 vs 1,180 bits).

\FloatBarrier
\section{Discussion}\label{sec:discussion}

The approach of this study departs from established practice in device forensics in two respects. First, in the drone and game console studies~\cite{clark2017}, \cite{eichhorn2024}, the meaning of the artifacts was determined by reverse engineering or by comparing images taken before and after pre-planned actions. The fact that Mayhem is open source made it possible to derive the same meaning directly from the source code and provided a basis for the interpretations in the absence of controlled action data. Second, the file-signature approach proposed for the Flipper Zero~\cite{fereli2025} cannot be applied to C16 captures, which have no identifying signature at the start of the file. Because Mayhem writes captures to contiguous clusters, the captures could be recovered without the need for a signature, from the free clusters between neighbouring captures. Unlike the study that uses an SDR as a signal-collection tool~\cite{binazhar2021}, here the device's own captures were decoded, and it was shown that the line-code guess of an automatic analyser can be misleading.

Table~\ref{tab:components} maps the components in Figure~\ref{fig:components} to the artifacts on the SD card. This mapping allows an examiner who holds the device to see which component produced which evidence. The SPI flash (Y7) and the clock ports (B2, B3) leave no direct trace on the SD card and must therefore be examined separately.

\begin{table}[!htb]
\centering
\caption{Relationship between hardware components and artifacts (codes as in Figure~\ref{fig:components}).}
\label{tab:components}
\footnotesize
\begin{tabularx}{\textwidth}{@{}P{3.6cm}P{4.2cm}L@{}}
\toprule
Component & Related artifact & Relationship \\
\midrule
B6 microSD slot & Entire image & The only persistent store of user data \\
K1 RTC backup battery & 1980-dated files, \texttt{DATE\_FILEFLAG} & Without a battery, the date starts from 1980 and advances by one day at every boot \\
B1 antenna port; Y3, Y4, Y5 radio circuits & \texttt{CAPTURES/*.C16}; frequency and gain values in \texttt{*.ini} & Captures are received through this port; \texttt{lna}, \texttt{vga}, \texttt{tx\_gain} are settings of these chips \\
E1 display & \texttt{SCREENSHOTS/*.PNG} & A screenshot is a copy of the frame buffer \\
D3, D4 controls & \texttt{SETTINGS/*.ini} & Every application exit writes an ini file \\
B4 micro-USB & \texttt{System Volume Information}, \texttt{.Trash-1000}, \texttt{gps-sdr-sim} & Computer access through SD Over USB or a card reader \\
Y7 SPI flash; D1, D2 buttons & \texttt{FIRMWARE/*.bin} & The actual location of the firmware is the flash; the copy on the SD card is indirect evidence \\
B2, B3 clock ports & \texttt{TX\_GPS.INI}, \texttt{TX\_ADSB.INI} & Applications sensitive to frequency accuracy; no trace of an external reference remains on the SD card \\
\bottomrule
\end{tabularx}
\end{table}

The findings lead to the following recommendations for practitioners:
\begin{enumerate}
  \item Remove the SD card before switching the device on. If an autostart application is configured, the device may transmit as soon as it is switched on; booting also changes \texttt{DATE\_FILEFLAG} and the settings files.
  \item Image the card through a write blocker, a device that prevents writing to the card, and copy the whole card as a physical disk rather than only the partition. In this way, the master boot record (MBR) at the start of the card and the area before the partition are also included in the copy.
  \item Keep the hash report of the acquisition tool together with the image; this report is the first proof that the image has not changed since acquisition.
  \item Record whether the RTC battery is present and the time shown on the screen; do not report 1980 dates as calendar dates (Section~\ref{sec:time}).
  \item Read settings files as evidence of use, not as evidence of transmission (Section~\ref{sec:catalogue}).
  \item Search for deleted IQ data in the free clusters around the capture directory (Section~\ref{sec:deleted}), and also read the firmware from the device with \texttt{hackrf\_spiflash -r}.
\end{enumerate}

The rules in Table~\ref{tab:catalogue} are based on the Mayhem v2.0.1 source code. Directory names and file formats have largely been preserved in later versions, but some applications have been turned into separate applications loaded from the SD card. The relevant source code should therefore be verified again for other versions. The timestamp, deletion, trimming and cluster-level recovery methods, on the other hand, can be applied directly to similar radio tools that use FAT32, for example the Flipper Zero~\cite{fereli2025}.

\FloatBarrier
\section{Limitations}\label{sec:limitations}

The study has the following limitations:
\begin{enumerate}
  \item The data come from unplanned, real use of the author's device. This approach provides realism, but it does not allow every artifact to be linked one-to-one to a known action. This shortcoming was partly addressed by deriving the interpretations from the source code, by internal consistency checks (Table~\ref{tab:validation}) and by the comparison with \texttt{rtl\_433} on sample captures. In addition, because the same person used and examined the device, the inferences have not been validated by an independent examiner; every inference was therefore based only on the traces in the image and on the source code.
  \item The findings are limited to Mayhem v2.0.1 and a single SD card.
  \item The Manchester decoder was tested only through its agreement with \texttt{rtl\_433}; no controlled capture of a transmitter with a known code was made.
  \item The SPI flash was not read. The firmware actually installed on the device and the serial number of the device could therefore not be verified directly.
  \item The conclusion that the device clock was set to local time is indirect; it rests on a comparison with the time at which the image was created.
  \item Because the image is a partition image, the MBR and the 1~MiB before the partition could not be examined.
  \item Before acquisition, the card was inserted into computers without a write blocker. The ``dirty'' flag at the start of the file system shows that the card was not ejected properly. It most probably remains from the Linux session on 2024-12-16, because the FatFs library used by Mayhem writes this sector only when formatting. The data overwritten during these connections could not be determined.
  \item The brand of the transmitter could not be determined in a search limited to the protocols recognized by \texttt{rtl\_433}.
\end{enumerate}

\FloatBarrier
\section{Conclusion}\label{sec:conclusion}

This study has identified, at source-code level, the artifacts that HackRF One/\allowbreak PortaPack devices write to the SD card under the Mayhem firmware and has proposed rules for their forensic interpretation. The core of the workflow consists of three operations: interpreting artifacts with a catalogue based on the source code, reading timestamps according to the writing system and the RTC state, and recovering deleted or trimmed IQ data and decoding them in a validated way. On the image taken from the author's own device, this workflow reached many conclusions from the traces in the image alone. The applications and sessions used, the date range of the period without a clock battery, and the trimming and editing performed on the device were determined; the final part of a deleted capture was recovered; and the captures were shown to belong to a single rolling-code transmitter. The two most important lessons for practitioners are to remove the SD card before the device is switched on and image it as a whole disk, and to keep the acquisition hash report together with the image.

The rules are limited to Mayhem v2.0.1 and a single SD card. The first step of future work is to carry out a differential analysis with controlled sets of actions on the same device and to compare the inferences drawn from the image with these actions. In addition, the rules should be validated on later versions of Mayhem, live acquisition through the SPI flash and the USB serial shell should be attempted, and an open-source parser that applies the artifact catalogue automatically should be developed.

In conclusion, this study offers an examiner who encounters a HackRF One/\allowbreak PortaPack a way, based on the source code and verifiable, to find, interpret and recover the usage traces on the SD card.

\section*{Data Availability}
\addcontentsline{toc}{section}{Data Availability}

The examined SD card image is publicly available on Zenodo under the CC BY 4.0 licence~\cite{hackrfimage}. It is distributed as an EnCase evidence file (E01) split into four segments. The SHA-256 value of the raw data it contains is \nolinkurl{99f6f02ef2e24df51566ad50280b421e6bae07d51a7e342d2e63d857ce2e948e}. All results reported in this paper can be reproduced from this image. The analysis scripts are available from the author on reasonable request. The Mayhem source code and the official release packages are publicly available~\cite{mayhem}.

\clearpage
\end{document}